# Curious Link of Exclusive and Inclusive CPV in Charmless 3-body B+ Decays

George W.-S. Hou

*Department of Physics, National Taiwan University,
Taipei 10617, Taiwan
E-mail: wshou@phys.ntu.edu.tw*

The LHCb experiment has measured *CP* violation (CPV) across the Dalitz plot of charmless decays of B+ mesons to 3 charged tracks, namely in K$\pi\pi$, KKK, $\pi\pi\pi$ and $\pi$KK final states, with strikingly large CPV that vary strongly with Dalitz variables. Identifying these processes with b → sq$\bar{q}$, ss$\bar{s}$ and b → dq$\bar{q}$, ds$\bar{s}$, where q = u, d, then the "sum rule" that requires two-loop absorptive parts by unitarity works well for inclusive b → s CPV, but less well for inclusive b → d case. We elucidate the situation, and argue that the 30 year old unitarity/*CPT* argument is valid to this day, while affirming quark-hadron duality.

## 1. The Striking Plot, and a Curiosity

The Bander-Silverman-Soni (BSS) mechanism[1] induces direct CPV (DCPV) in B+ decay via the b → sq$\bar{q}$ transition: CPV phase in $V_{ub}$ via the Tree (T) diagram, the Penguin (P) diagram gives absorptive part via b → si$\bar{i}$ (i = u, c) then on-shell i$\bar{i}$ → q$\bar{q}$ rescattering. Since the B factory discovery of charmless 3-body B decays, the LHCb experiment has now measured inclusive CPV for B±, namely[2]

$$A_{CP}(B^\pm \to K^\pm\pi^+\pi^-) = +0.025 \pm 0.004 \pm 0.004 \pm 0.007 \text{ [PDG: +0.027]}, \quad (1)$$

$$A_{CP}(B^\pm \to K^\pm K^+K^-) = -0.036 \pm 0.004 \pm 0.002 \pm 0.007 \text{ [PDG: -0.033]}, \quad (2)$$

$$A_{CP}(B^\pm \to \pi^\pm\pi^+\pi^-) = +0.058 \pm 0.008 \pm 0.009 \pm 0.007 \text{ [PDG: +0.057]}, \quad (3)$$

$$A_{CP}(B^\pm \to \pi^\pm K^+K^-) = -0.123 \pm 0.017 \pm 0.012 \pm 0.007 \text{ [PDG: -0.122]}, \quad (4)$$

and across the Dalitz plot, e.g. in Fig. 1 for B± → $\pi^\pm\pi^+\pi^-$, the "artist's palette". Strikingly large CPV vary strongly across the Dalitz plot, reflecting the presence of hadron level rescattering and resonant interference. The asymmetries in Fig. 1 are in bins with same number of events, where one has[2] roughly 181k and 109k events in K$\pi\pi$ and KKK modes,[a] and 25k and 6k in $\pi\pi\pi$ and $\pi$KK.

---

[a] Note that LHCb has excellent K/$\pi$ separation or PID, while pions suffer more background.

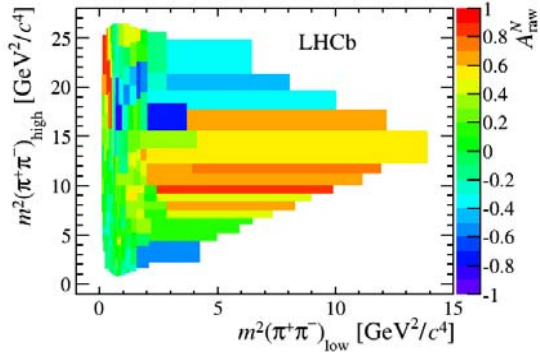

Fig. 1: CPV across the Dalitz plot[2] for $B^\pm \to \pi^\pm\pi^+\pi^-$ decay, an "artist's palette".

The curiosity is then, when weighted by PDG values of ~ 5.1, 3.4, 1.5, 0.5 ($\times 10^{-5}$) for the respective decay branching ratios, and asymmetries given in [...] in the equations without error, combining Eqs. (1) and (2) one finds 0.3%, and combining Eqs. (3) and (4) one finds 1.1%, for CPV in b → s and b → d transitions, respectively. From large DCPV approaching $\pm O(1)$ locally in the Dalitz plot, to a few % to O(10%) for the inclusive charmless mesonic 3-body final states, the DCPV for inclusive charmless $B^\pm$ decay to even or odd number of kaons in the final state is actually sub-percent! This is reminiscent of a "prediction" made 30 years ago,[3,4] where quark level inclusive b → s$q\bar{q}$ and b → d$q\bar{q}$ decays were found at the sub-percent level (a "sum rule"), when taking into account an additional T-P interference, with extra absorptive part from off-shell gluon propagator in P as demanded by unitarity and *CPT* arguments.

In the following, since *CP*-invariant phase is needed for CPV, we first discuss Final State Interactions (FSI), progressing from elastic to inelastic, and from soft (hadronic) to hard, such as the annihilation of $c\bar{c}$ into charmless $q\bar{q}$ (q = u, d, s) from the leading b → sc$\bar{c}$ decay, crucial for the penguin process.

## 2. FSI: elastic vs inelastic

In summer 2001, Belle observed[5,b] the color-suppressed $\bar{B}^0 \to D^0\pi^0$ decay at 9.3σ, and with the $D^{*0}\pi^0$, $D^0\eta^0$ and $D^0\omega^0$ modes above 4σ. The observed rates were much larger than theory predictions of the time, indicating the presence of FSI,[6] or elastic $D^{(*)}\pi \to D^{(*)}\pi$ rescattering, while the $D^{(*)0}\eta$ and $D^{(*)0}\omega$ final states imply SU(3) symmetry. Before long, the formalism was extended[7] to the more lucrative charmless two body B decays, such as B → K$\pi$, to pursue DCPV. This now involved meson loops, with D or light mesons running in the loop, which raises the questions: How does it work when the mesons in the loop are far off-shell? What does one do for inelastic scattering vertices when far off-shell hadrons are involved? The diagramatics and discussions, followed to this day,[8] are really just cherry-picking, lack in control mechanism, and quite arbitrary. It is an approach that we have purposely stayed away from.

---

[b] If half a year later, the leading authors would have been K.-F. Chen, H.-C. Huang and R.-S. Lu. This is one of two first discovery papers by the young NTU-Belle group.

But we are certainly not adverse to FSI. Pre-B-factory, in the CLEO era, we had already pursued[9,10,11] FSI vigorously, for sake of the phases it may provide that are necessary for DCPV. And when elastic FSI was experimentally observed in $B^0 \to D^0\pi^0$, we extended the work[6] to suggest something similar in $B \to K\pi$, i.e. experimental indications for presence of elastic $K\pi \to K\pi$ FSI scattering.[12] This work was actually quite fine, but was not viewed favorably by PRL reviewers. But the consolation soon came with Belle observation[13,c] of DCPV in $B^0 \to K^+\pi^-$ decay, which was again the effort of NTU-Belle.

We note that the original expectation in the late 1990s of QCD factorization in charmless two body B decays did not come to pass. For ourselves, when the experimentally firm DCPV difference[14] between $B^0 \to K^+\pi^-$ and $B^+ \to K^+\pi^0$ was not corroborated by a large and negative $\sin\phi_s$, suggesting instead "enhanced color-suppressed" amplitude C as culprit, we no longer trusted perturbative QCD estimates of strong phases in B decay to hadronic final states.

### 3. FSI: soft vs hard

In a 2015 LHCb talk[15] presented in Brazil (and in other LHCb papers), attempt was given to elucidate the volatility of DCPV across the Dalitz plot via $\pi^+\pi^- \leftrightarrow K^+K^-$ rescattering. Invoking *CPT* invariance, it was assumed that "all the inelasticity of the $\pi\pi$ interaction goes into the KK channel" in the rescattering region of $(1, 2.2)$ GeV$^2$, which was rooted in an earlier discussion[16] that "*CPT* symmetry constrains hadron rescattering so that the sum of the partial decay widths of all channels <u>with the same final-state quantum numbers related by the scattering matrix</u> must equal that of their charge-conjugated decays".[2] We scrutinize this notion: Although the statement is not incorrect, the underlined part (by us) is often not sufficiently respected, i.e. treated in some cavalier way. We then broaden our criticism before turning to quark level language.

### 3.1. *CPT and inelastic hadron scattering*

The upper plots[15] in Fig. 2 zooms in on the CPV in Dalitz plot of lower $\pi^+\pi^-$, $K^+K^-$ mass regions for $B^\pm \to K^\pm\pi^+\pi^-$, $K^\pm K^+K^-$ decays, where the positive tendency for $B^\pm \to K^\pm\pi^+\pi^-$ seems to counterbalance the negative tendency for $B^\pm$

---

→ $K^{\pm}K^{+}K^{-}$ in the $m^2_{\pi\pi}$, $m^2_{KK}$ regions of (1, 2.2) GeV$^2$ (as marked), lending credence to $\pi^{+}\pi^{-} \leftrightarrow K^{+}K^{-}$ rescattering[2,15,16] as argued by *CPT*. However, the scattering matrix would connect $\pi^{+}\pi^{-} \to K^{+}K^{-} + (n\pi)^0$ that have "the same final-state quantum numbers" for n = 0–3, where $(n\pi)^0$ is charge neutral. These thresholds can be drawn for $B^{\pm} \to K^{\pm}\pi^{+}\pi^{-}$ at $m^2_{\pi\pi} \approx$ 1.28, 1.62, 1.99 GeV$^2$, respectively, and much denser for $K^{+}K^{-} \to$ multi-pion final states for $B^{\pm} \to K^{\pm}K^{+}K^{-}$. The scattering matrix is thus rather large and inelastic, and thinking only in terms of $\pi^{+}\pi^{-} \leftrightarrow K^{+}K^{-}$ rescattering as guided by *CPT* is too simplistic.

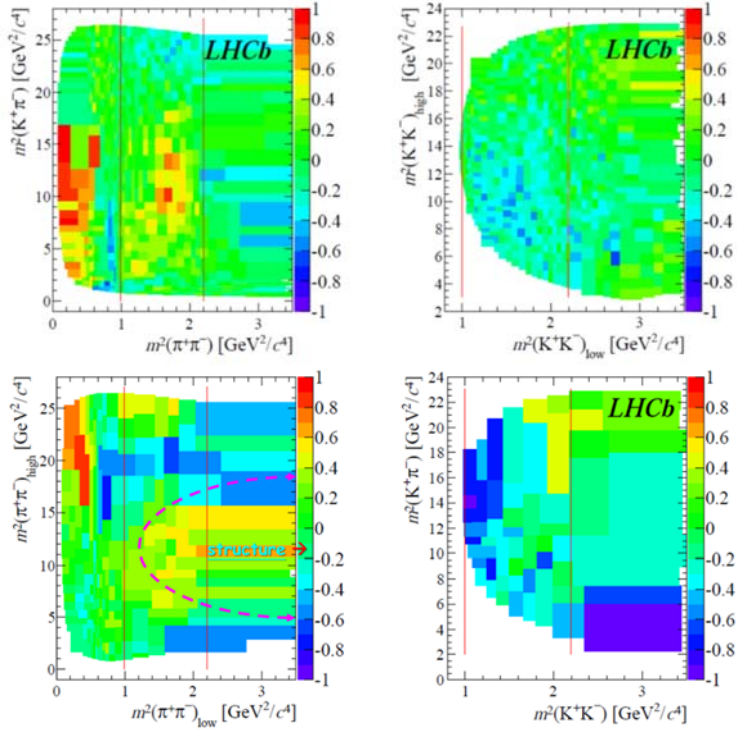

Fig. 2: CPV in Dalitz plot for [upper] $B^{\pm} \to K^{\pm}\pi^{+}\pi^{-}$, $K^{\pm}K^{+}K^{-}$ and [lower] $B^{\pm} \to \pi^{\pm}\pi^{+}\pi^{-}$, $\pi^{\pm}K^{+}K^{-}$ decays, with (1, 2.2) GeV$^2$ as marked.[15]

A similar argument applied to $B^{\pm} \to \pi^{\pm}\pi^{+}\pi^{-}$, $\pi^{\pm}K^{+}K^{-}$ decays as illustrated in lower plots of Fig. 2 makes the case further, where even the LHCb discussion[15] admits to "other mechanisms in action" for the $\pi^{\pm}\pi^{+}\pi^{-}$ final state.

### 3.2. *a Critique*

The lower-left plot in Fig. 2 actually zooms into the lower $m^2_{\pi\pi}$ region of the "artist's palette", the most colorful Fig. 1 for $B^{\pm} \to \pi^{\pm}\pi^{+}\pi^{-}$ decay, where the latter gives a broader picture. LHCb has recently conducted an amplitude analysis[17] of the upper-left corner of rather low mass $m_{\pi\pi} \sim m_\rho$, together with the high mass $m_{\pi\pi}$ above 4 GeV or so. The analysis went beyond S-P wave interference,[2,15] and observed DCPV also in D-wave, but things are still not fully settled. Note that we have marked in the lower-left plot of Fig. 2, that there is some starting "structure" towards the right as outlined by the (magenta) dashed curve, which extends into the brilliant and bright yellow-to-red region of the full Dalitz plot of Fig. 1. The lower boundary of this structure is around the $c\bar{c}$ threshold, namely an underlying c-anti-c to q-anti-q annihilation effect, where q is a light quark (u, d, s). This $c\bar{c} \to q\bar{q}$ annihilation amounts to $D\bar{D} + X \to$ charmless rescattering,[4] as stated in the original BSS mechanism at quark level.

We see that there are "soft" FSI, which amounts to $K\pi\pi \leftrightarrow KKK$ and $\pi\pi\pi \leftrightarrow \pi KK$ rescattering, where an $n\pi$ system can be thrown in easily on either side to add inelasticity. The complexity seems rather difficult to address. But one also has hard FSI, where $c\bar{c} \to q\bar{q}$, or equivalently a double charm (but with no overall charm quantum number — *CPT* connection!) $D\bar{D} + X$ rescattering into 3-body charmless mesonic final states from the leading $B \to D\bar{D} + X_s$ decay process, as discussed in our work 30 years ago, which brings about subtleties.[3,4]

We now turn to discuss *CPT* and unitarity at the quark level.

## 4. Unitarity/CPT at Quark Level: a 30-year "Sum Rule"

In 1988 we published a study[18] of inclusive $b \to sg^*$ decay, where the $g^*$ could be light-like (on-shell gluon), time-like ($g^*$ splits to $q\bar{q}$,[d] which is the usual Penguin for B decays), and space-like (the gluon attaches to the spectator). We corrected the $b \to sg$ formula and showed that the time-like process dominates, in contrast to the usual Penguin for kaon decay, where one cannot distinguish between time-like and space-like processes, while light-like is ill-defined.

The quark level $b \to sq\bar{q}$ process would match to inclusive charmless $B \to K + X$ decays by quark-hadron duality, where X has no net strange quantum number. Having done the inclusive rate, and being a student of Amarjit Soni, it was natural to explore inclusive CPV via the BSS mechanism.[1] At MPI Munich, we paired up with Jean-Marc Gérard, dividing efforts between inclusive vs exclusive processes. We soon ran into a paradox: the direct application of BSS mechanism did not seem to respect unitarity, nor *CPT*. We will not go into the details of the discussion[3,4] 30 years ago (including a debate with Lincoln Wolfenstein[19]), but just give a brief account and come back to the comparison with charmless $B^+$ decays to 3-body mesonic final states measured by LHCb.

---

[d] The treatment[3,18] of $g^* \to gg$ was later shown to be absent (see PRD article[4] for references).

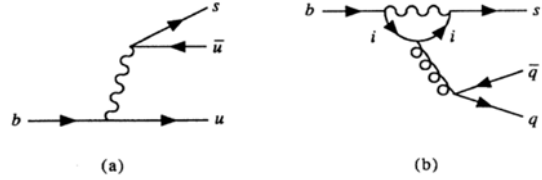

Fig. 3: BSS mechanism[1] for DCPV in b → sq$\bar{q}$ decay (taken from Ref. 4) via T-P interference.

### 4.1. *Unitarity/CPT at Quark Level*

Fig. 3 corresponds to the BSS mechanism[1] of T-P interference, where CPV phase is in T amplitude, while the absorptive part arises from the i = c cut of the Penguin loop. Although the mechanism was intended for T-P interference such as for b → su$\bar{u}$ process, with the pure Penguin process at hand, we purposed to study CPV in pure Penguin process, where one can imagine P-P interference, with one dispersive P-amplitude providing CPV phase, while the other P-amplitude is absorptive. But we soon found that this procedure for T-P interference in b → su$\bar{u}$ process, extended to include P-P interference for pure penguin b →su$\bar{u}$, sd$\bar{d}$ and ss$\bar{s}$ processes, did not respect unitarity, nor *CPT*. The situation was the same for the corresponding b → du$\bar{u}$, dd$\bar{d}$ and ds$\bar{s}$ processes. We refer to the PRD article[4] for a detailed exposition.

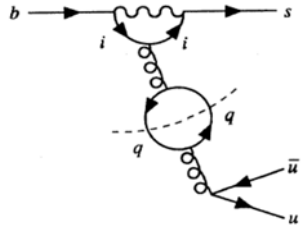

Fig. 4: Diagram (taken from Ref. 4) needed to interfere with Fig. 3(a) to restore unitarity/*CPT*.

It turns out that, to maintain unitarity and *CPT*, one needs to include an additional two-loop diagram that interferes with Fig. 3(a): the "double-Penguin" displayed in Fig. 4! The gluon propagator carries a "bubble" or q$\bar{q}$ loop (the q = u, d, s cut illustrated by dashed line) which possesses a rather large absorptive part (see footnote d). Interfering with only b → su$\bar{u}$ of Fig. 3(a), it tends to cancel the CPV in this mode and restore unitarity/*CPT*.

One can visualize this unitarity/*CPT* argument. In Fig. 5 we make[4] a term-by-term amplitude-squared analysis of the BSS mechanism. Fig. 5(a) is T-T interference in b → su$\bar{u}$, which conserves *CP* because there is no "strong phase", while one has the CKM factor $|V_{ub}|^2$ hence no CPV phase. Fig. 5(b) is the usual BSS mechanism, or T-P interference. However, as explained, we purposed to study DCPV in pure penguins, which is Fig. 5(c), where the i = u and c cuts of P-P interference illustrates two possibilities for quark level strong phases.[e] It was found that, stopping at this level, where pure Penguin decays such as b → sdd̄, ss̄s would also possess DCPV, one would not respect unitarity nor *CPT* (see PRD article [4] for detailed discussion). One needs to include Fig. 5(d).

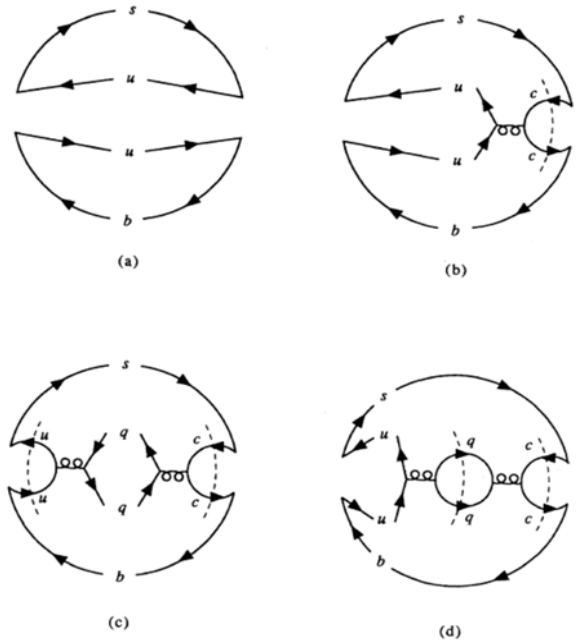

Fig. 5: (a) T-T interference (b → suū); (b) T-P interference (b → suū), or BSS mechanism; (c) P-P interference (pure Penguin b → sqq̄); (d) T–double-P interference. Figure taken from Ref. 4.

Comparing Figs. 5(d) and 5(c), one can understand how they are connected. The amplitude-squared in Fig. 5(c) also represents the b → sqq̄ decay rate (q = u, d, s) to charmless final state. But for the u-quark cut in the Penguin, it could also correspond to the b → suū process illustrated in Fig. 5(d), with the q = u, d, s final states now correspond to the cut of the Penguin bubble, or absorptive part of a time-like gluon propagator. The two diagrams are therefore related by unitarity (the "same" diagram), hence ignoring Fig. 5(d) amounts to a fallacy. As for *CPT*, note that on-shell color-octet qq̄ and cc̄ have <u>the same final-state</u>

---

[e] The dispersive top-loop provides a subtraction to the u- and c-loops by GIM mechanism.

quantum numbers related by the scattering matrix, via one gluon exchange. At the quark level, there are no issues as discussed in Sec. 3 with hadronic processes. Thus, these final states feed each other,[4] governed by *CPT*. Such arguments guided us 30 years ago to identify the interference of T (Fig. 3(a)) and "double-P" (Fig. 4) that restored unitarity/*CPT*.

TABLE I. (Semi-)inclusive branching ratio $B$ and asymmetries for $b \to s$ and $b \to d$ processes for $\rho = -0.5$, $\eta = 0.15$, and $m_t = M_W$. $a_0$ is with Figs. 1(a) and 1(b) only, while $a$ is the result with Fig. 1(c) taken into account. The entry "0.0" stands for a very small positive number.

|  | $B$ (%) | $a_0$ (%) | $a$ (%) |
|---|---|---|---|
| $b \to s u \bar{u}$ | 0.46 | 1.2 | 0.0 |
| $b \to s d \bar{d} + s s \bar{s}$ | 0.54 | 0.5 | 0.5 |
| Total $b \to s$ (no charm) | 1.19 | 0.7 | 0.2 |
| $b \to d u \bar{u}$ | 0.71 | −0.7 | −0.0 |
| $b \to d d \bar{d} + d s \bar{s}$ | 0.07 | −4.2 | −4.2 |
| Total $b \to d$ (no charm) | 0.80 | −1.0 | −0.4 |

Fig. 6: Table I, or numerical results, from Ref. 4. Note $\rho \approx -0.12$, $\eta \approx 0.35$ at present.

### 4.2. *a 30-year "Sum Rule"*

To illustrate the numerics, let us first display Table I from the PRD article[4] in Fig. 6. This corresponds to $\rho = -0.5$, $\eta = 0.15$, ca. 1990, for the Wolfenstein parameters governing $V_{ub}$. The current values are $\rho \approx -0.12$, $\eta \approx 0.35$ are rather different, so we are only concerned with gross features.[f] But by including Fig. 4, i.e. from $a_0$ to $a$ in Fig. 6 (Table I), the asymmetries for $b \to su\bar{u}$ and $b \to du\bar{u}$ drop to 0, while the inclusive $b \to sq\bar{q}$ asymmetry drop from percent level (noting that $\eta$ is now 2x larger) to 0.2%. This is what we mentioned as a "prediction" from almost 30 years ago. We will comment on $b \to dq\bar{q}$ later.

It is heuristic to match $b \to sq\bar{q}$ to charmless $B^\pm$ decay to odd number of kaons in 3-body mesonic final state. But one may wish to match $b \to su\bar{u}$ with $B^\pm \to K^\pm \pi^+ \pi^-$, and $b \to ss\bar{s}$ with $B^\pm \to K^\pm K^+ K^-$, which may also seem plausible at first sight. However, if one takes $B^\pm \to K^\pm$ as matching to $b \to s$, then $[K^-\pi^+\pi^-]$ and $[K^-K^+K^-]$ can be represented as $[s \bar{u}d \bar{u}d]$ and $[s \bar{u}s \bar{u}s]$, respectively. Thus, both $b \to su\bar{u}$ and $b \to sd\bar{d}$ feed $B^\pm \to K^\pm \pi^+ \pi^-$, while both $b \to ss\bar{s}$ and $b \to su\bar{u}$ feed $B^\pm \to K^\pm K^+ K^-$ (though $b \to ss\bar{s}$ should dominate over $b \to su\bar{u}$), so the matching of quark level processes to the mesonic level ones are not trivial. Besides the wide difference in ρ, η values ca. 1990, this makes the comparison of Eqs. (1) and (2) with b → su$\bar{u}$ and b → sd$\bar{d}$ + ss$\bar{s}$ entries difficult.

---

[f] Given the numerical accuracy, we do not distinguish between $\rho$, $\eta$ and the "barred" counterparts.

With the heuristic argument that b → ds$\bar{s}$ should dominate over b → du$\bar{u}$ in generating $B^\pm \to \pi^\pm K^+K^-$, and that η is now a little more than 2x larger than ca. 1990, it is intriguing that ~ −12% from Eq. (4) seems to match ~ −4% of Table I quite well. What is troubling is the +1.1% asymmetry for inclusive $B^\pm$ decay to even number of kaons in 3-body mesonic final state, which does not match the negative sign in Table I. But ρ ≈ −0.5 ca. 1990 is rather different from the current value of −0.12. The trouble is, to update the 30 year old result, one needs to fully update the effective field theory, including the on-shell rescattering as outlined. We shy away from it at this stage given the current level of precision. This also implies that further experimental updates are desired, to eliminate statistical fluctuations as another possible cause of disagreement.

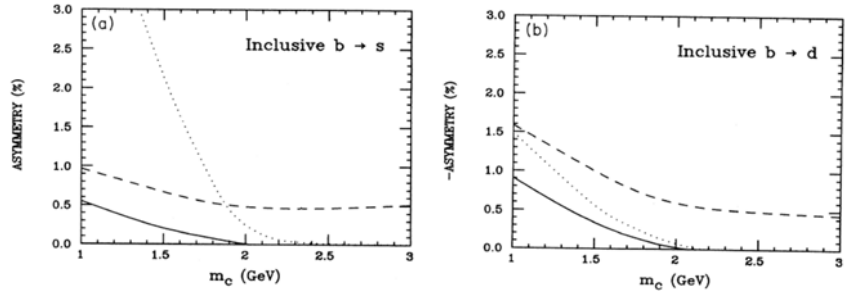

Fig. 7: Inclusive (a) b → s and (b) b → d asymmetries vs $m_c$,[4] where dotted line is Fig. 5(b) alone, dashed line includes Fig. 5(c), and solid line includes Fig. 5(d). *CPT* requires inclusive asymmetry to vanish when $2m_c > m_b$,[3,4] which is not respected by the dashed line, but restored when Fig. 5(d) is included. The smallness of the inclusive asymmetry is in part due to c$\bar{c}$ threshold suppression.

Before commenting further on how to update the theory, we address a question of curiosity: if pure Penguin DCPV, as well as the T−double-P interference are at $O(\alpha_s^2)$ hence higher order, why is it numerically so important? The issue turns out to be the relative proximity of the c$\bar{c}$ threshold to $m_b$, and the factor of 3 from q = u, d, s for pure Penguin final state. In Fig. 7 we give the inclusive b → s and b → d asymmetries from our PRD article.[4] As explained in the caption, the inclusion of T−double-P interference, Fig. 5(d), restores *CPT*, and one sees threshold suppression as behind the smallness of the inclusive asymmetry. But at the deeper level, if the c$\bar{c}$ cut on the P-amplitude suffers threshold suppression, the double-Penguin, where the time-like gluon absorptive part brings in q$\bar{q}$ cut which does not suffer threshold suppression, it brings in further a factor of 3 enhancement due to three light quarks. Thus, the impact of $O(\alpha_s^2)$ process over $O(\alpha_s)$ process for charmless B decays is indeed a subtle one. It is quite satisfying that it is now reflected in charmless 3-body mesonic decays measured by LHCb.

Of course, one should keep in mind that the smallness of $|V_{ub}|$ elevates the importance of the Penguin.

### 4.3. *Improving the Inclusive DCPV Theory*

As stated, we have not attempted to update the numerics from 30 years ago.

- Interested theorists can try to improve things, but it's a little muddied and complicated, as one needs to update to an effective field theory language, not just updating ρ, η and $m_b$, $m_c$, as we have already commented.
- One needs to take into account the light-like penguin b → sg, with g on-shell. This process has been updated[20,21] and estimated at O(0.5%), hence not small compared with time-like penguin b → sq$\bar{q}$. But "on-shell" gluons do not give much absorptive parts, so how much does it contribute to DCPV? Further, in what part of "h′±h−h+" (h = K, π) Dalitz plot does b → sg interfere with b → sq$\bar{q}$? These would not be easy questions to address, and would have to be incorporated in the effective field theory.
- The good news is that the near cancellation of inclusive asymmetries is guaranteed by the strength of the absorptive part of the double-Penguin. But does it hold true for ρ that is considerably weaker than what was considered around 1990? And, is there any way to access the individual quark level processes such as b → su$\bar{u}$, sd$\bar{d}$, ss$\bar{s}$ separately? In any case, it would be interesting to see further experimental (and theoretical) development.

### 5. Conclusion

Thirty years ago,[3,4] "*we found the surprising result that $O(\alpha_s^2)$ contributions are as important as $O(\alpha_s)$ effects for individual semi-inclusive modes such as* b → su$\bar{u}$. *These unexpected results were uncovered because we chose to study the rates and asymmetries of semi-inclusive charmless* B *decays, and we therefore had to pay better attention to general conditions such as unitarity and CPT.*" And "*it is indeed subtle*".[4]

> "*The upshot of our results is that the total inclusive charmless* b → s *and* b → d *decay asymmetries are rather suppressed.*" [4]

In 2014, the LHCb experiment measured[2] DCPV in charmless $B^{\pm}$ → h′±h−h+ decays, with rather large asymmetries that vary strongly across the Dalitz plot. Such hadronic level asymmetries are not precluded by the above statement, but data does support sub-percent CPV in charmless inclusive 3-body decays, especially for $B^+$ decay to an odd number of charged kaons, corresponding to charmless inclusive b →

s$q\bar q$. This affirms the rather small inclusive charmless asymmetry, or "sum rule", stated above, and also affirms quark-hadron duality holds well in such decays. For charmless 3-body $B^+$ decay to an even number of charged kaons that correspond to charmless inclusive b → dq$\bar q$, the inclusive asymmetry seems just above percent level, which should be further clarified.

**Acknowledgments**

This work is supported by grants MOST 106-2112-M-002-015-MY3 and NTU 108L104019.